\documentclass[aps, prb, reprint, showpacs]{revtex4-1}
\usepackage{amsmath, amsfonts, amssymb, bm}
\usepackage{color, graphicx}
\usepackage{multirow}
\usepackage[colorlinks,citecolor=blue, urlcolor=blue,
  linkcolor=blue]{hyperref}
\begin{document}

\title{Kondo lattice model: from local to non-local descriptions} 

\author{Gang Li}
\email{gangli@physik.uni-wuerzburg.de}
\affiliation{\mbox{Institut f\"ur Theoretische Physik und Astrophysik,
  Universit\"at W\"urzburg, 97074 W\"urzburg, Germany} }

\begin{abstract}
In this paper, we study the influence of spatial fluctuations in a
two-dimentional Kondo-Lattice model (KLM) with anti-ferromagnetic
couplings. 
To accomplish this, we first present an implementation of the
dual-fermion (DF) approach based on the hybridization expansion
continuous-time quantum Monte Carlo impurity solver (CT-HYB), 
which allows us to consistently compare the local and non-local
descriptions of this model.
We find that, the inclusion of non-locality restores the
self-energy dispersion of the conduction electrons, {\it i.e.} the
$\vec{k}$ dependence of $\Sigma(\vec{k}, i\omega_{n})$.
The anti-ferromagnetic correlations result in an additional symmetry in
$\Sigma(\vec{k}, i\omega_{n})$, which is well described by the
N\'eel antiferromagnetic wave-vector. 
A ``metal"-``anti-ferromagnetic insulator"-``Kondo insulator"
transition is observed at finite temperatures, which is driven by the
competition of the effective RKKY interaction (at the weak coupling
regime) and the Kondo singlet formation mechanism (at the strong
coupling regime).   
Away from half-filling, the anti-ferromagnetic phase becomes unstable
against hole doping. The system tends to develop a ferromagnetic phase
with the spin susceptibility $\chi_{s}(Q)$ peaking at $Q=\Gamma$.
However, for small $J/t$, no divergence of $\chi_{s}(\Gamma)$ is
really observed, thus, we find no sign of long-range ferromagnetism
in the hole-doped two-dimension KLM.
The ferromagnetism is found to be stable at larger $J/t$ regime. 
Interestingly, we find the local approximation employed in this work,
{\it i.e.} the dynamical mean-field theory (DMFT), is still a very
good description of the KLM, especially in the hole-doped case.  
We do not observe clear difference of the two-particle spin
susceptibilities in the DMFT and the DF calculations in this regime. 
However, at half-filling, the non-local fluctuation effect is indeed
pronounced. 
We observe a strong reduction of the critical coupling strength for
the onset of the Kondo insulating phase.
\end{abstract}


\maketitle

\section{Introduction}
{\it Heavy fermion}, as the name tells, is a fermion with large
effective mass as compared to that of a non-interacting fermion. 
They are found in a large number of lanthanide and actinide
compounds~\cite{PhysRevLett.35.1779, Fulde19881, nuki1987281,
  Steglich, RevModPhys.56.755}, which display many striking phenomena.  
The large effective mass can be seen, for example, from the specific
heat of Ce$_{x}$La$_{1-x}$Cu$_{6}$~\cite{nuki1987281}, which shows a
large linear temperature-dependence.  
The effective mass is known to be propositional to this linear
coefficient in Fermi liquid theory, whose quasiparticle description is
believed to be valid in the heavy fermion systems~\cite{PhysRevLett.60.1570}.
The specific heat seems to have a universal doping dependence, which
indicates that each Ce ion independently contributes to the enhanced
specific heat, thus the ion-electron interaction is rather local. 
The measured temperature-dependence of the uniform susceptibility
shows two different types of behaviors~\cite{PhysRevLett.35.1779,
  nuki1987281}. 
A Curie-Weiss behavior is found at temperatures higher than a characteristic
temperature $T_{K}$, which is known as the kondo temperature.
The susceptibility is non-monotonic at  low T, and eventually saturates to a finite value when $T$ approaches 
to zero with the value being a few orders of magnitude larger than
values typical for simple metals. This leads to the enhanced Pauli
susceptibility for $T<T_{K}$.

To account for the local coupling between the magnetic moments and
itinerant electrons, the Kondo lattice model (KLM) is proposed,
\begin{equation}\label{H_kondo}
  H = -t\sum_{\langle
    i,j\rangle,\sigma}(c_{i,\sigma}^{\dagger}c_{j\sigma} + h.c.) - 
  \mu\sum_{i,\sigma}c_{i\sigma}^{\dagger}c_{i\sigma} + 
  J\sum_{i}\vec{S}_{i}^{f}\cdot\vec{s}_{i}^{c}.
\end{equation}
where $c_{i,\sigma}^{\dagger}(c_{i\sigma})$ creates(annihillates) of a
conduction electron with spin $\sigma$ at the $i$-th site. 
Eq.~(\ref{H_kondo}) describes localized degrees of freedom interacting
with itinerant electrons. It is an effective model for many materials
with f orbitals (some with d orbitals), in which the Coulomb
interaction between electrons are so strong that the charge
fluctuations are essentially suppressed, leading to a situation that
only spin/orbital degree of freedom remain active. 
This process can be mathematically simulated by cutting off the one-electron hybridization
processes in the periodic Anderson model as done in the
Schrieffer-Wolff transformation.~\cite{PhysRev.149.491}
In Eq.~(\ref{H_kondo}), at each site, the spin sector of the conduction
electrons interacts locally with that of the f-orbital. 
This interaction gives rise to the dominant energy scale in the large $J/t$
limit, where the local magnetic impurity is fully screened by the
conduction electron by forming spin singlets, {\it i.e.} Kondo
effect.
However, going from the PAM to the KLM, the adiabatic connection to
$J/t=0$ limit is lost. 
Unlike in the PAM, the perturbation series with respect to $J/t$ is
singular at $J=0$~\cite{PhysRevLett.102.017202}, which makes the
solution of the KLM non-trivial even for small $J/t$.
When $J/t$ is small, another energy scale emerges from the coupling of
the conduction electron polarization around the magnetic impurities at
different sites, {\it i.e.} the so-called Ruderman-Kittel-Kasuya-Yosida
(RKKY) interaction.
The RKKY interaction is an indirect interaction of the local magnetic
impurities induced by Kondo exchange coupling. 
To understand the competition of these two energy scales is one of the
key issues in the study of heavy fermion systems.

Theoretically, the phase diagram of the KLM has been extensively
studied by the mean-field theory~\cite{PhysRevB.66.045111,
  PhysRevB.20.1969, 10.1007} 
and numerical approaches in one-dimension~\cite{PhysRevLett.65.3177,
  PhysRevB.44.7486, PhysRevLett.71.3866}. 
For reviews of the KLM, see, for example,
references~\onlinecite{RevModPhys.69.809, ColemanReview, Kuramoto}.
Recently, many efforts~\cite{PhysRevLett.102.017202,
  PhysRevB.81.113108, JPSJ.78.034719, JPSJ.78.014702,
  PhysRevB.87.165133} are dedicated to the study of the KLM in the
infinite-dimension by using the dynamical mean-field theory 
(DMFT)~\cite{RevModPhys.68.13}. 
The DMFT is an exact mapping of the interacting many-body problem to
an effective impurity problem in the infinite-dimension, in which
spatial fluctuations are completely neglected. 
In this paper, we study the anti-ferromagnetic Kondo model on a
two-dimensional square lattice, with special attention to the
non-local fluctuation effect. 
Effort of going beyond the local description of the KLM has been
recently carried out~\cite{PhysRevB.82.245105,
  PhysRevLett.101.066404} at two-dimension by using one
cluster-extension of the DMFT, {\it i.e.} the dynamical  
cluster approximation (DCA)~\cite{RevModPhys.77.1027}. 
In the DCA, the short-range correlations inside a two-site cluster was fully taken into
account. 
Here, we take a different strategy, we consider the
non-local corrections to the DMFT local solution via the local
two-particle vertices through the dual fermion
approach~\cite{PhysRevB.77.033101, PhysRevB.79.045133}, in which both
the short- and long-range non-local fluctuations can be treated on
equal footing, but they both are only approximately taken in account
in this method.
We calculate the one-body self-energy function $\Sigma(\vec{k},
i\omega_{n})$ and identify the non-local fluctuations from the
momentum dependence of it. 
We find that the spatial fluctuation is pronounced at half-filling and
becomes weaker with hole doping. 
At the hole doped regime, we find no sign of long-range ferromagnetism
at small $J/t$. 
The ferromagnetic spin arrangement is only stabilized when $J/t$ is
large. 
We find the DMFT solution is quite close to the DF results
with large hole doping, indicating spatial fluctuation is negligible
in this case.   
To see the non-local effect on the anti-ferromagnetic phase, we
calculate $\chi_{s}^{DFMT}(Q,\Omega_{m}=0)$ from the DMFT and
$\chi_{s}^{DF}(Q, \Omega_{m}=0)$ from the DF approach.  
We find the non-local fluctuations suppresses the antiferromagnetic
phase to lower temperature and smaller $J/t$ regime.

This paper is organized as follows: we start with the local
description of the KLM by solving the DMFT equation with the
CT-HYB impurity solver.  
We briefly summarize the basic idea of the CT-HYB for the Kondo
problem in~\ref{Local_Alg}. In~\ref{Local_Res} we present a few DMFT
results focusing on the two-dimension case. A comparison to the
infinite-dimension DMFT results is made, which also serves as
benchmarks of our implementation. 
In section~\ref{Non_Local}, the non-local extension of the DMFT is
given. The details of the DF scheme for the Kondo lattice
model is presented in~\ref{NL_Alg} and the main discussion of
this paper is dedicated to the non-local fluctuations on the single-
and two-particle properties, which is presented
in~\ref{NL_Res}. 
Summary and outlook are then given in section~\ref{Con}. 

\section{LOCAL DESCRIPTION}\label{Local}

\subsection{algorithm}\label{Local_Alg}
In the language of the DMFT, the many-body
interacting problem in Eq.~(\ref{H_kondo}) is locally approximated by
a magnetic impurity embedded in a continuous bath of the conduction
electrons. 
The non-local degrees of freedom in Eq.~(\ref{H_kondo}), {\it i.e.}
the kinetic term $-t\sum_{\langle
  i,j\rangle,\sigma}(c_{i,\sigma}^{\dagger}c_{j\sigma} + h.c.)$, can
be readily integrated out.  
Due to the coupling of the $c-$ and the $f-$electrons at the impurity
site, integrating over the bath mathematically results in a dynamic
function $\Delta(\omega)$ associated to the impurity, which
accounts for the hybridization of the impurity with the
itinerate conduction electrons. 
 The local Hamiltonian is given as:
\begin{equation}\label{H_local}
H_{loc}^{i} = -\mu\sum_{\sigma}c_{i\sigma}^{\dagger}c_{i\sigma} +
J\vec{S}_{i}^{f}\cdot\vec{s}_{i}^{c}.
\end{equation}

Eq.~(\ref{H_local}) can be easily diagonalized in
the particle-number basis. 
Table~\ref{Eigen} shows the corresponding eigenfunctions and eigenvalues.
The full Hilbert space of $H_{loc}^{i}$ is a product of those of the
conduction electrons and the magnetic impurity. As for a spin-$1/2$
impurity which is what we consider in this work.
The Hilbert space dimension is 8.  
Hamiltonian Eq.~(\ref{H_local}) conserves particle number $N$ and spin
SU(2) symmetry, thus the corresponding Hilbert space can be
further decomposed into 7 blocks characterized by different quantum
numbers $(N, S_{z}^{tot})$. As one can see, the largest dimension of
these decoupled blocks is 2.
The decompsion of the full Hilbert space is of great convenience for the
numerical simulation of this model that we will specify below. 
\begin{table}[htbp]
  \centering
  \begin{tabular}{|c|c|c|c|}
\hline
\hline
    Energy & Eigenstates & Particle Number & $S_{z}^{tot}$\\ 
\hline
    0    &   $|1\rangle=|0, \downarrow\rangle$  & 0 & -1/2 \\ 
\hline
    0    &   $|2\rangle=|0, \uparrow\rangle$  & 0 & 1/2 \\ 
\hline
$J/4-\mu$   & $|3\rangle=|\downarrow, \downarrow\rangle$  & 1 & -1 \\ 
\hline
$-3J/4-\mu$ & $|4\rangle=\frac{1}{\sqrt{2}}(|\uparrow,
\downarrow\rangle-|\downarrow,\uparrow\rangle)$  &\multirow{2}{*}{1} &
\multirow{2}{*}{0} \\  
\cline{1-2}
$J/4-\mu$ & $|5\rangle=\frac{1}{\sqrt{2}}(|\uparrow,
\downarrow\rangle+|\downarrow,\uparrow\rangle)$  &  &  \\      
\hline
$J/4-\mu$   & $|6\rangle=|\uparrow, \uparrow\rangle$  & 1 & 1 \\ 
\hline
$-2\mu$   & $|7\rangle=|\uparrow\downarrow, \downarrow\rangle$  & 2 & -1/2 \\ 
\hline
$-2\mu$   & $|8\rangle=|\uparrow\downarrow, \uparrow\rangle$  & 2 & 1/2 \\ 
\hline
  \end{tabular}
  \caption{Eigenvalues and eigenstates of the local Hamilotnian
    Eq.~(\ref{H_local}) in the particle-number basis. The Hilbert
    space is further decoupled into different blocks with respect to
    the total particle number and the $z$-component of the total spin
    operator. The first arrow in each eigenstate represents the
    $c$-electron spin, while the second arrow represents the spin of
    the local magnetic impurity.} 
  \label{Eigen}
\end{table}

\begin{figure}[htbp]
\centering
\includegraphics[width=\linewidth]{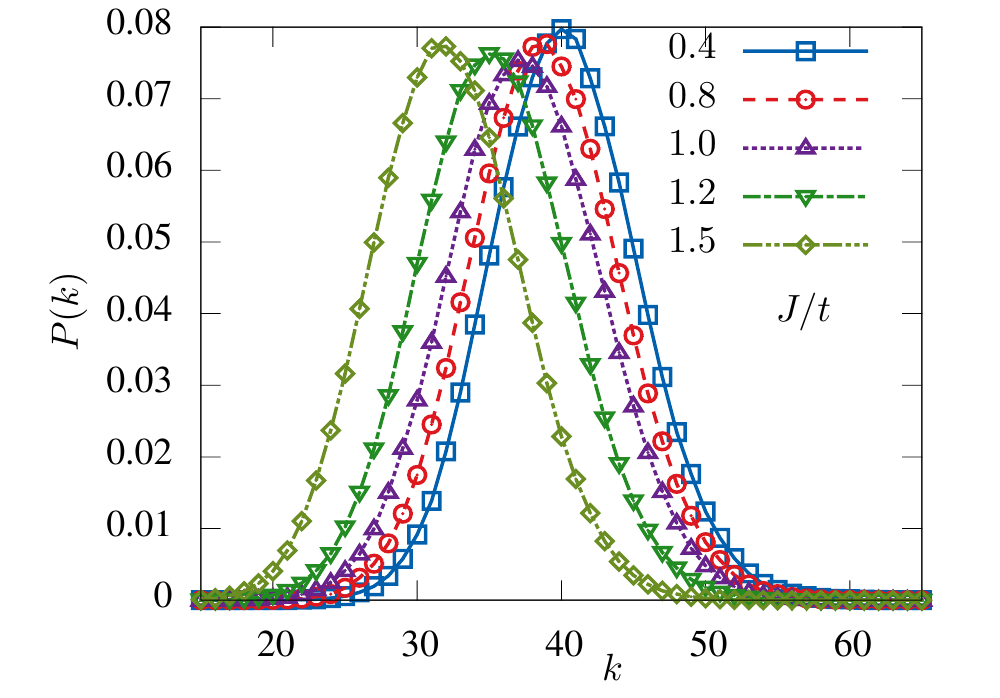}
\caption{The  probability of the expansion terms in
  Eq.~(\ref{H_kondo}) being sampled at each order for the Kondo
  lattice model on a two-dimension square lattice at $\beta t=50$.} 
\label{Pk}
\end{figure}
The complete DMFT action of the Kondo lattice model is given as the
sum of the hybridization and the local parts.
\begin{eqnarray}\label{dmft}
S[\{c\}; \{f\}] &=&
-\sum_{\sigma}c_{\sigma,\omega_{n}}^{*}[i\omega_{n}+\mu -
  \Delta(i\omega_{n})]c_{\sigma,\omega_{n}} \nonumber \\
&& +J\vec{S}^{f}\cdot\vec{s}^{c} 
\end{eqnarray}
In this paper, to numerically solve this action, we employ the
continuous-time quantum Monte Carlo method~\cite{RevModPhys.83.349}
based on the hybridization-expansion
algorithm~\cite{PhysRevLett.97.076405, PhysRevB.74.155107}.
We note that, the interaction-expansion algorithm~\cite{PhysRevB.72.035122}
has also been applied to a closely related model, {\it i.e.}
Coqblin-Schrieffer model with N components~\cite{PhysRev.185.847} by
J. Otsuki {\it et al.}~\cite{JPSJ.78.014702, JPSJ.78.034719}.
Here, we employ the implementation recently proposed by us~\cite{PhysRevB.85.115103} directly to the KLM.
 In this implementation, the single- and two-particle Green's functions are
directly sampled in the Matsubara frequencies space, no imaginary-time
measurement is required.  
The noise of the self-energy function at large frequencies are
removed by replacing the self-energy with a carefully determined
high-frequency tail.   
This tail is calculated independently from the CT-HYB simulation. We find
that, usually, only a few numbers of Matsubara frequencies need to be 
simulated. And this number decreases with the increase of interaction
strengths.  
Thus, moderate speedup of the simulation can be gained in this
implementation. 

As for the Kondo problem defined in Eq.~(\ref{dmft}), the basic idea of
the CT-HYB algorithm can be understood as an expansion over the
coupling between the conduction electrons and the magnetic impurity. 
\begin{eqnarray}\label{expansion}
  {\cal Z}_{imp}&& = \mbox{Tr}e^{-\int_{0}^{\beta}\mbox{d}\tau\mbox{d}\tau^{\prime}
  S(\tau,\tau^{\prime})} = {\cal Z}_{c}{\cal Z}_{loc} 
  \sum_{k}\frac{1}{k!^{2}}\mbox{Tr}\times\nonumber \\
  &&\langle T_{\tau}\int_{0}^{\beta}\{\mbox{d}\tau\}
  c(\tau_{1})c^{\dagger}(\tau_{1}^{\prime})\cdots
  c(\tau_{k})c^{\dagger}(\tau_{k}^{\prime})\rangle_{c}
  \langle Det^{{\cal C}_{k}}\rangle_{b}, 
\end{eqnarray}
where $k$ is the expansion order, ${\cal
  Z}_{c}=\mbox{Tr}_{c}\exp[-\beta H_{c}]$, ${\cal 
  Z}_{loc}=\mbox{Tr}_{c}\exp[-\beta H_{loc}]$ are the partition
function corresponding to the conduction electrons and the local
Hamiltonian.  
If we choose the basis function of ${\cal Z}_{loc}$ as
the eigenfunctions listed in table~\ref{Eigen}, ${\cal Z}_{loc}$ can
be easily calculated from the eigenenergies of $H_{loc}$.  
However, under this basis, $c(\tau)$ and $c(\tau^{\prime})^{\dagger}$
in Eq.~(\ref{expansion}) become matrices, whose elements connect
different eigenstates. 
Due to the conservation of the quantum numbers mentioned in
table~\ref{Eigen}, these elements are non-vanishing only between
certain eigenstates.
Thus, the matrix product of the list of kinks, {\it e.g.} $c(\tau)$
and $c^{\dagger}(\tau^{\prime})$ in Eq.~(\ref{expansion}), only need to be
calculated between certain blocks of the full Hilbert space.
Compared to the production of matrix of size $8\times8$ for the full Hilbert space, now the largest matrix needs to be treated is $2\times1$. Thus, the decomposition greatly reduces the simulation efforts.
In Eq.~(\ref{expansion}), every expansion term will be faithfully
calculated in a stochastic way in the simulation.  
Therefore, the CT-HYB can provide a numerically exact solution to the
DMFT mapping of the Kondo lattice model.  
Under the MC importance sampling algorithm, for our problem, there is
only finite number of terms have non-vanishing contributions to the
expansion.  

\subsection{results}\label{Local_Res}

The DMFT + CT-HYB study of the Kondo lattice model has been carried
out on the bethe lattice~\cite{PhysRevB.74.155107}.  
The study presented in this section focuses on the square lattice
geometry, we will mainly discuss the influence of the system-dimension
reduction. 
In addition, they can also be viewed as benchmark of our
implementations. 

\begin{figure}[htbp]
\centering
\includegraphics[width=\linewidth]{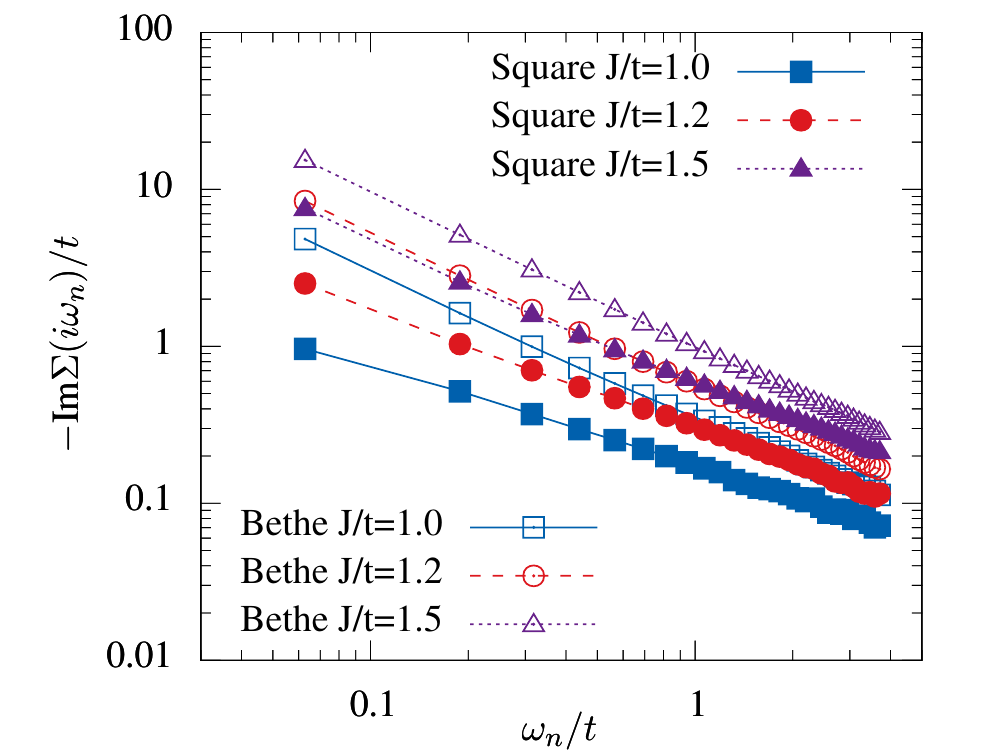}
\caption{The comparison of the local self-energy of the
  antiferromagnetic KLM at $\beta t=50$ on a
  two-dimension square and an infinite-dimension bethe lattice. }  
\label{bs}
\end{figure}
Fig.~\ref{Pk} shows an estimate of the expansion order "k" (see Eq.~(\ref{expansion})) for
different values of $J/t$ on the square lattice at $\beta t=50$. As
on the bethe lattice~\cite{PhysRevB.74.155107}, the increase of $J/t$
value leads to the movement of the distribution $P(k)$  towards a
smaller value of $k$.
This is because the stochastic expansion in the CT-HYB is around the
atomic limit, the larger coupling $J/t$ would justify the local
Hamiltonian Eq.~(\ref{H_local}) to be a better zero approximation of
the full Hamiltonian, {\it e.g.} Eq.~(\ref{dmft}).
$P(k)$ can be a measure of the numerical
expense of the problem~\cite{PhysRevLett.97.076405}, it defines the
dimension of the determinantal matrix and the number of ``kinks" 
to be multiplied~\cite{PhysRevB.74.155107}.
Fig.~\ref{Pk} shows that the larger values of $J/t$ cases can be more
efficiently simulated in the CT-HYB algorithm. 
Compared to $P(k)$ for the bethe lattice
calculations~\cite{PhysRevB.74.155107}, we find that for given value
of $J/t$, the reduction of system dimension from infinity- to
two-dimension shifts the distribution to larger value of $J/t$ (, for the
corresponding distribution $P(k)$ on the bethe lattice,
see~\onlinecite{PhysRevB.74.155107}). 

As another benchmark of our simulations, we show in Fig.~\ref{bs} the
local self-energy and compare them to the counterparts on the bethe
lattice~\cite{PhysRevB.74.155107}.  
The solid symbols in Fig.~\ref{bs} represent the imaginary part of the
local self-energy for $J/t=1.0, 1.2$ and $1.5$ on the square lattice. 
The empty symbols correspond to the solutions on the bethe lattice with
the same parameters.  
The reduction of system dimension in the DMFT does not change the
self-energy too much, though $P(k)$ shown in Fig.~\ref{Pk} centers at ``k" of value more than
two times larger on the square lattice than on the bethe lattice.   
$\Sigma(i\omega_{n})$ behaves very similarly for the two different
lattice geometries, {\it i.e.} $-\mbox{Im}\Sigma(i\omega_{n})/t$
logarithmly increases with the decrease of frequencies $\omega_{n}$,
which suggests the system to be an insulator.  

\begin{figure}[htbp]
\centering
\includegraphics[width=\linewidth]{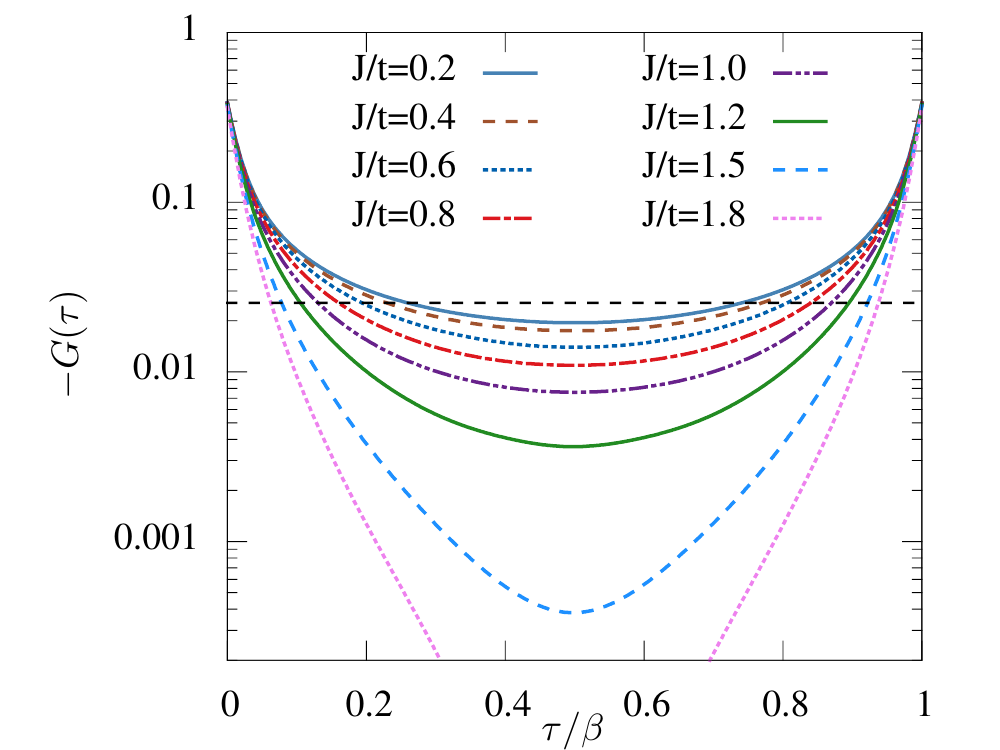}
\caption{Imaginary-time Green's function of the conduction electrons
  at inverse temperature $\beta t=50$. With the increase of coupling
  strengths, $-G(\tau)$ at $t=\beta/2$ exponentially decays. But for
  all values of $J/t$, $-G(\beta/2)$ stay below the value (the black
  dashed line) expected for the Fermi liquid.} 
\label{Gt}
\end{figure}
We can further confirm this conclusion by examining the imaginary-time
Green's function of the conduction electrons in Fig.~\ref{Gt}.
We can clearly see that for all values of $J/t$, $G(\beta/2)$
remains smaller than $4/(\pi\beta t)\approx0.02546$ expected for a
Fermi liquid.
$G(\beta/2)$ varies very slowly with decreasing of $J/t$, thus, we
could reasonably expect that for any small value of $J/t$ the system
will never become a Fermi liquid. 
Further decreasing of temperatures would lead to a smaller value of
$G(\beta/2)$ for given $J/t$.
This leads to the conclusion that, he ground state of the anti-ferromagnetic Kondo model may be an insulator for arbitrary value of $J/t$. 

\begin{figure}[htbp]
\centering
\includegraphics[width=\linewidth]{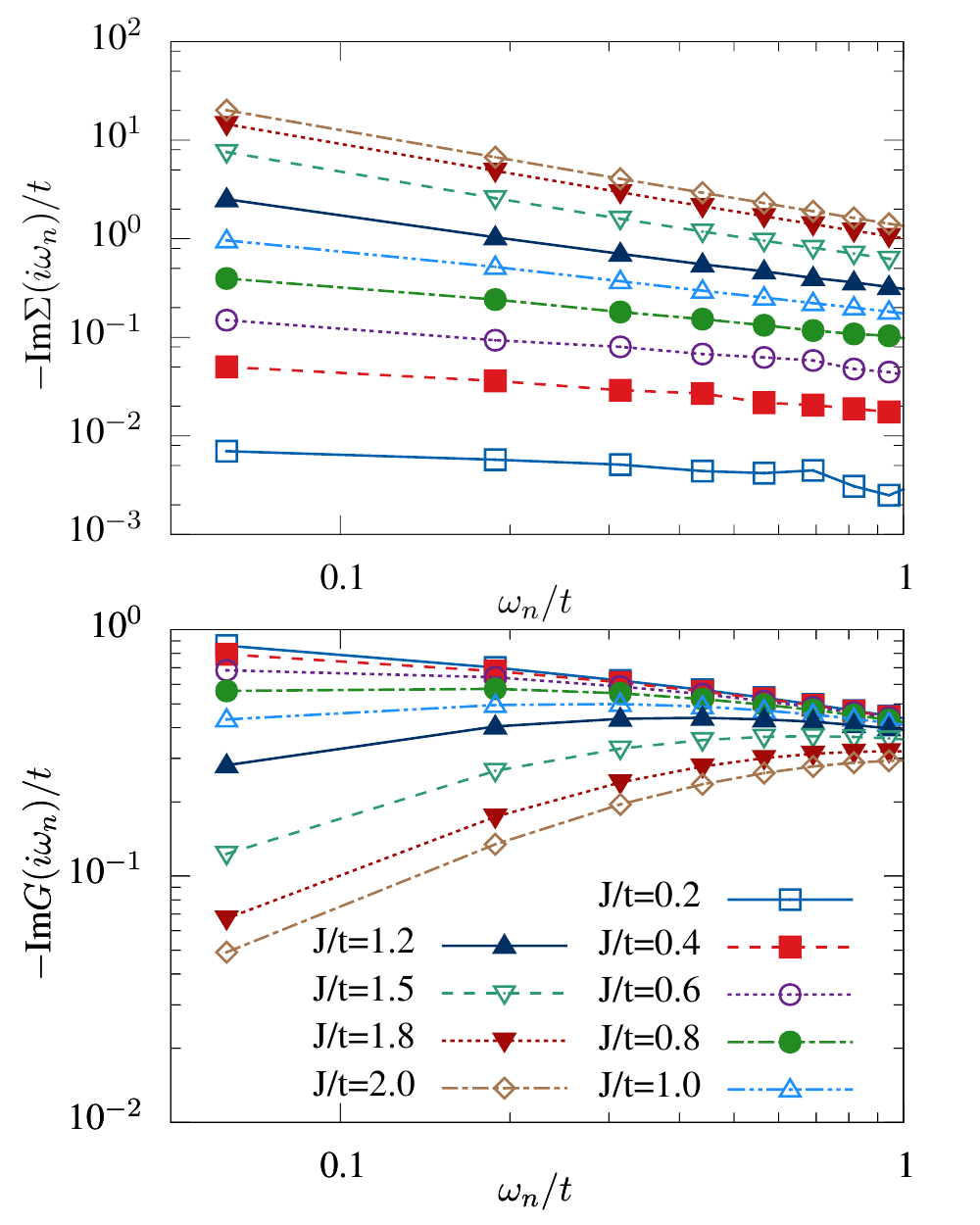}
\caption{The imaginary part of the self-energy function and the
  corresponding one-particle Green's function of the conduction
  electrons. In the upper panel, $\Sigma(i\omega_{n})$ exponentially
  increases when $\omega_{n}$ approaches to zero. 
  In the lower panel, $G(i\omega_{n})$ shows no divergence
  and approaches to zero for large values of $J/t$ at
$\omega_{n}\rightarrow0$. }
\label{SE}
\end{figure}
Fig.~\ref{SE} displays the imaginary part of the self-energy function
for $\beta t=50$ and various $J/t$.
Compared to Fig.~\ref{bs}, here results for more values of $J/t$ are
presented and they all correspond to the square lattice study. 
$-\mbox{Im}\Sigma(i\omega_{n})/t$ logarithmly increases with the
decrease of $\omega_{n}$ for all values of $J/t$. 
Decreasing $J/t$ only slowly increase the slopes of
$-\mbox{Im}\Sigma(i\omega_{n})/t$, but never changes the sign of the
slope to positive values.  
Thus, the self-energy will never fall down to zero at
$\omega_{n}\rightarrow0$ as expected for the metallic phase.
This leads to an insulating phase for all values of $J/t$, which
agrees with the behavior of $G(\beta/2)$.
However, the one-body Green's function shown in the lower plot of
Fig.~\ref{SE} seems to display different behaviors at larger and smaller
values of $J/t$.
In general, the single-particle Green's function is divergent in the
metallic phase and approaches to zero in the insulating phase as
$\omega_{n}\rightarrow0$. 
As shown in Fig.~\ref{SE}, for $J/t$ larger than $0.8$,
$-\mbox{Im}G(i\omega_{n})t$ remains finite (not divergent) at
$\omega_{n}\rightarrow0$. 
It becomes flat at $J/t=0.8$ and starts to decrease with the further
increase of $J/t$ as expected for insulators.
From Fig.~\ref{Gt} and Fig.~\ref{SE}(a), we know the system is
insulating for all values $J/t$ studied here. 
The non-vanishing value of $-\mbox{Im}G(i\omega_{n})t$ for $J/t>0.8$
at $\omega_{n}\rightarrow0$ in Fig.~\ref{SE}(b) reflects that the
charge gap is very small in these cases, which is essentially the same
the situation as of $G(\beta/2)$ for smaller values of interaction,
where they are very close to the value expected for a Fermi liquid
(see the dashed line in Fig.~\ref{Gt}).

\begin{figure}[htbp]
\centering
\includegraphics[width=\linewidth]{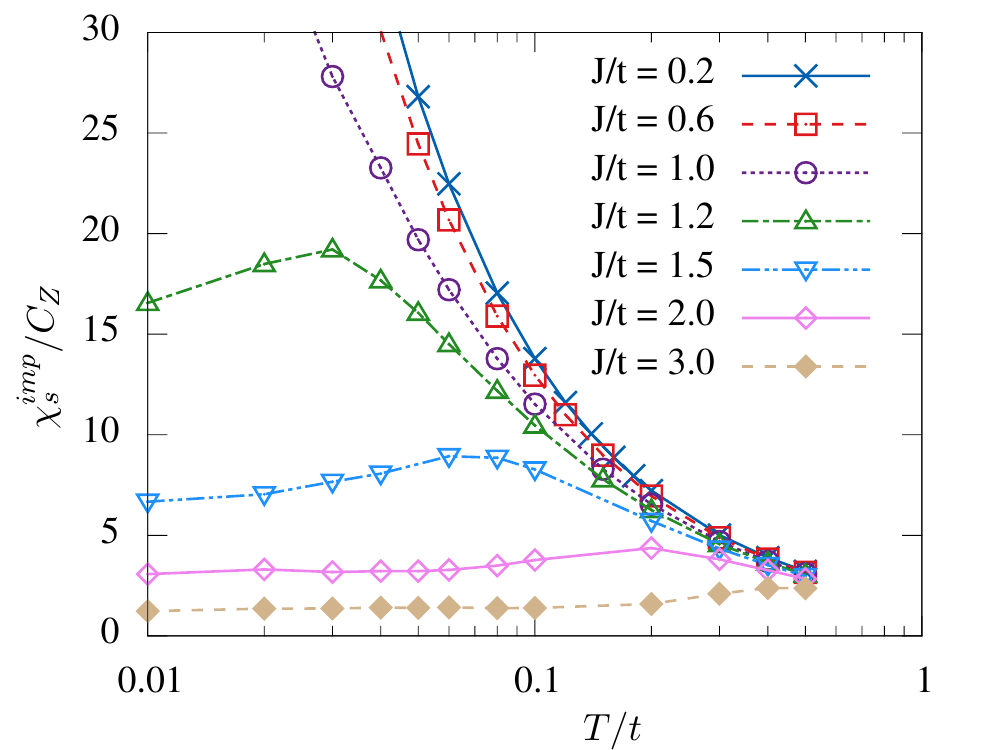}
\caption{The uniform spin susceptibility of the KLM, which contains
  the contribution from both local moments and conduction
  electrons. $J\chi_{s}(T)$ shows a crossover from $\sim 1/T$ at high temperatures to
  $\sim 1/T_{K}$ behaviors at low temperatures.}
\label{chi}
\end{figure}
To track the crossover from high temperature magnetic moments to low
temperature electrons screening processes, we show in Fig.~\ref{chi}
the temperature evolution of the impurity spin susceptibility
\begin{equation}
\chi_{s}^{imp}(T) = \int_{0}^{\beta}d\tau \chi^{imp}_{s}(\tau,
T)=\int_{0}^{\beta}d\tau\langle S_{z}(\tau)S_{z}(0)\rangle
\end{equation}
and normalize it with the static susceptibility $C_{Z}=\langle
S_{z}^{2}\rangle$. 
In the CT-HYB, both $\chi_{s}^{imp}(T)$ and $C_{Z}$ can be measured to
very high precision. 
In the eigenbasis shown in Table~\ref{Eigen}, $S_{z}$ is simply a
conserved number, which is the z-component of the total spin of
Eq.~(\ref{H_kondo}). 
Thus, $\chi_{s}^{imp}(T)$ actually contains the contribution from both local
moments and conduction electrons.
This is different from the quantities shown in Fig.~\ref{Gt} and
Fig.~\ref{SE}, which only correspond to the conduction electrons.
Due to the conservation of $S_{z}$, in the CT-HYB, the matrix product
in Eq.~(\ref{expansion}) with $S_{z}(\tau)$ and $S_{z}(0)$ operators
can be trivially done.
As we expected, $\chi_{s}^{imp}(T)$ shows two distinct behaviors at high and
low temperatures.
At high temperature $\chi_{s}^{imp}(T)$ scales in Curie law $\sim 1/T$,
while at low temperature it saturates and shows a paramagnetic
behavior $\sim 1/T_{K}$.
The temperature $T_{K}$ marks the crossover of the susceptibility from
a high temperature Curie-Weiss type to a low temperature Pauli type.
With the increase of $J/t$, the saturation of $\chi_{s}^{imp}(T)/C_{Z}$
moves to larger value of $T/t$, indicating that the Kondo temperature
$T_{K}$ increases with the increase of the Kondo coupling strengths. 
Following reference~\onlinecite{JPSJ.78.034719}, one can estimate the Kondo
temperature $T_{K}$ from the inverse of the impurity spin susceptibility as
$T_{K}=C_{Z}/\chi_{s}^{imp}$ at sufficiently low temperature. 
Corresponding to the parameters in Fig.~\ref{chi}, $T_{K}$ can be
found in Fig.~\ref{Mag}, which we will discuss later on together with
the magnetic phases detected from the DMFT and the dual-fermion 
approaches, respectively. 

\section{NON-LOCAL CORRECTION}\label{Non_Local}

To go beyond the local approximation of the DMFT, we consider the
non-local corrections generated by the local two-particle vertices,
through the dual-fermion approach~\cite{PhysRevB.77.033101,
  PhysRevB.79.045133}. 
In the DMFT mapping of the Kondo problem defined on the square
lattice, the one-particle hopping of the conduction electrons
is essentially replaced by the hybridization function
$\Delta(\omega)$, which locally couples to the impurity. 
Such a replacement is exact when the dimension of the system is
infinite. 
For the problem defined on the square lattice, $\epsilon(k)$ cannot be
exactly mapped to $\Delta(\omega)$. 
As the DMFT is a good approximation in many cases,
$\epsilon(k)$ - $\Delta(\omega)$ can be small. 
Thus, we treat $\epsilon(k)-\Delta(\omega)$ as a small parameter
and perturbatively study their influence on the DMFT solution. 
This is the basic idea of the dual-fermion approach.
As a comprehensive presentation of our implementation of the
dual-fermion method in conjunction with the CT-HYB, we explain the
methodology of the dual-fermion approach in details here once again,
following the original work of
A.N.~Rubtsov and his collaborators.~\cite{PhysRevB.77.033101}

\subsection{algorithm}\label{NL_Alg} 
Mathematically, one can formulate the above idea by rewriting the
lattice action with that of the impurity and expands
$\epsilon(k)-\Delta(\omega)$ around the DMFT limit. 
Under the path-integral, the lattice partition function is expressed
as 
${\cal Z}=\int{\cal D}[\{c\};\{f\}]e^{-S}$, and
\begin{eqnarray}\label{lattice}
S[\{c\};\{f\}]&=&-\sum_{k,\sigma}c^{*}_{k,\omega_{n},\sigma}[i\omega_{n}+\mu
  -\epsilon_{k}]c_{k,\omega_{n},\sigma}\nonumber \\
&& +J\sum_{i}\vec{S}^{f}\cdot\vec{s}^{c}.
\end{eqnarray}
We can add/subtract $\Delta(i\omega_{n})$ to/from the above action,
which essentially changes nothing, but now the action reads
\begin{equation}
S[\{c\};\{f\}]=\sum_{i}S_{imp}^{i} +
\sum_{\alpha}c^{*}_{\alpha}
    [\epsilon_{k}-\Delta(i\omega_{n})] c_{\alpha},
\end{equation}
where $S_{imp}$ is given by Eq.~(\ref{dmft}) and the short notation
$\alpha=(k,\omega_{n},\sigma)$ is used.
As stated above, if we take $\epsilon_{k}-\Delta(i\omega_{n})$ as an
expansion parameter (no matter it is large or small), we can formally
expand this action around $S_{imp}$. 
If we are able to collect every term, this expansion will still be
an exact expression of $S[\{c\};\{f\}]$.
In the dual-fermion approach, this expansion is practically carried
out by applying one of the path-integral standard technique, {\it
  i.e.} change of variable.

The last term in the above expression can be rewritten in an equivalent
form by introducing a new set of variables, {\it e.g.} $\{d\}$.
\begin{eqnarray}
e^{-\sum_{\alpha}c_{\alpha}^{\dagger}A_{\alpha}c_{\alpha}}
=\det^{-1}A
\int{\cal D}[d,d^{*}]e^{(-c^{*}_{\alpha}d_{\alpha}+h.c.) 
  - d^{*}_{\alpha}A_{\alpha}d_{\alpha}},
\end{eqnarray}
which depends on both the conduction electron degrees of freedom, {\it
  i.e.} $\{c\}$, and the introduced dual variables, {\it i.e.} $\{d\}$. 
Here $A$ denotes matrix
$\widehat{A_{k\omega_{n},\sigma}}$, with element
$A_{k\omega_{n},\sigma}=[\Delta(i\omega_{n})-\epsilon_{k}]^{-1}$. 
The partition function of the lattice problem defined in
Eq.~(\ref{H_kondo}) now depends on all three variables,
\begin{eqnarray}\label{new-action}
{\cal Z} &=&\det^{-1}A\int 
{\cal D}[c,c^{*};f,f^{*}]e^{-\sum_{i}S_{imp}^{i}}\times\nonumber\\
&&\int {\cal
  D}[d,d^{*}]e^{-\sum_{\alpha}[c_{\alpha}^{*}d_{\alpha} + d_{\alpha}^{*}c_{\alpha} 
  +d_{\alpha}^{*}A_{\alpha}d_{\alpha}]}\phantom{.} 
\end{eqnarray}

We should note here, there is no approximation involved in the above
derivations, Eq.~(\ref{new-action}) is an equivalent expression of the lattice
action.  
Thus, any quantity of interests can be equally evaluated through
Eq.~(\ref{new-action}) and Eq.~(\ref{lattice}).
For example, one can calculate the single-particle Green's function
from both actions (by taking $\epsilon_{k}$ as the source term and
  differentiate both actions over it), which leads to
\begin{equation}\label{G-fd}
G_{k\omega_{n},\sigma}=[\Delta(i\omega_{n})-\epsilon_{k}]^{-2}G^{dual}_{k\omega_{n},\sigma}
+ [\Delta(i\omega_{n})-\epsilon_{k}]^{-1}, 
\end{equation}
where $G^{dual}_{k\omega_{n},\sigma}=-\langle
d_{k\omega_{n},\sigma}d^{*}_{k\omega_{n},\sigma}\rangle$
denotes the single-particle Green's function of the dual variables. 

It becomes transparent now that the introduction of the new variable changes
the calculation of the lattice Green's function $G_{k\omega_{n},\sigma}$ to
that of $G^{dual}_{k\omega_{n},\sigma}$.
With respect to the complexity of solving the KLM, it
seems that nothing is achieved in the transformation in Eqs.~(\ref{lattice}
- \ref{G-fd}), as $G^{dual}_{k\omega_{n},\sigma}$ is not known and its
calculation can be equally complicated as that of $G_{k\omega_{n},\sigma}$.
However, as one will see below, due to the fact that the expansion is
around the DMFT solution, $G_{k\omega_{n},\sigma}^{dual}$ can be reliably
calculated in a perturbative manner, which is much simpler than the
calculation of $G_{k\omega_{n},\sigma}$ directly.

As for the Kondo problem we study in this paper, the evaluation of
$G^{dual}_{k\omega_{n},\sigma}$ formally requires an action that
depends only on $\{d\}$ and $\{f\}$ degrees of freedom, which can
be obtained by integrating $\{c\}$ out of Eq.~(\ref{new-action}).
$\{c\}$ and $\{d\}$ are separated in Eq.~(\ref{new-action}) except for $ 
(c_{k\omega_{n},\sigma}^{*}d_{k\omega_{n},\sigma}+h.c.)$. 
Expanding the partition function in Eq.~(\ref{new-action}) over
this mixed term, and neglecting any term in which $c$ and $c^{*}$ are
not paired (according to the Grassmann algebra) leads to 
\begin{widetext}
\begin{eqnarray}\label{dual-Z}
{\cal Z} &=&\det^{-1}A\int
     {\cal D}[d, d^{*}] 
     \exp(-\sum_{\alpha}d^{*}_{\alpha}A_{\alpha}d_{\alpha})\times{\cal Z}_{imp}\int{\cal
       D}[c,c^{*};f,f^{*}]e^{-\sum_{i}S_{imp}^{i}[c,c^{*};f,f^{*}]}\times\nonumber\\
     &&\times[1+\sum_{\alpha_{1}\alpha_{2}}d^{*}_{\alpha_{1}}\langle
       c_{\alpha_{1}}c_{\alpha_{2}}^{*}
       \rangle_{imp}d_{\alpha_{2}} 
     +\frac{1}{4}\sum_{\alpha_{1}\alpha_{2}\alpha_{3}\alpha_{4}}\langle
       c_{\alpha_{1}}c_{\alpha_{2}}^{*}c_{\alpha_{3}}c_{\alpha_{4}}^{*}\rangle_{imp} 
     d_{\alpha_{1}}^{*}d_{\alpha_{2}}d_{\alpha_{3}}^{*}d_{\alpha_{4}}
     + \mbox{higher orders}] \nonumber\\
     &=&{\cal Z}_{imp}\det^{-1}A\int{\cal D}[d, d^{*}] 
     \exp(-\sum_{\alpha}d^{*}_{\alpha}(A_{\alpha}+g_{\alpha})d_{\alpha} - V[d,d^{*}]),
\end{eqnarray}
\end{widetext}
where $g_{\alpha}=-\langle c_{\alpha}c_{\alpha}^{*}\rangle_{imp}$ and
$\chi_{1234}=\langle
c_{\alpha_{1}}c_{\alpha_{2}}^{*}c_{\alpha_{3}}c_{\alpha_{4}}^{*}\rangle_{imp}$
are the impurity one- and two-particle Green's functions.
A complete separation of vaiables $\{d\}$ with the local degress of
freedom $\{c\}$ and $\{f\}$ is achieved in this equation.
In the last step, we have brought each expansion term back to the exponential
function and grouped all terms of order higher than 1 to $V[d,d^{*}]$. 
The first term in $V[d, d^{*}]$ is 
$\frac{1}{4}\gamma_{1234}d_{\alpha_{1}}^{*}d_{\alpha_{2}}d_{\alpha_{3}}^{*}d_{\alpha_{4}}$,
with $\gamma_{1234}$ the reducible part of $\chi_{1234}$ (, the bubble
term is subtracted due to the exponential form of
$\sum_{\alpha_{1}\alpha_{2}}d^{*}_{\alpha_{1}}\langle c_{\alpha_{1}}c_{\alpha_{2}}^{*} 
       \rangle_{imp}d_{\alpha_{2}}$). 

$g_{\alpha}$ and $\gamma_{1234}$ are given as the solutions of the DMFT
calculations. 
As a technical remark, we note that, though with the CT-HYB as imprity
solver, the higher-frequency part of the Matsubara self-energy
function $\Sigma_{\alpha}$ contains larger statistical error than the
lower-frequency part, the sampling of $g_{\alpha}$ and $\gamma_{1234}$ is more
stable and less affected by the statical noises.
Moreover, due to the subtraction of the bubble contribution from
$\chi_{1234}$, $\gamma_{1234}$ can be reliably sampled directly in the
Matsubara frequency space~\cite{PhysRevB.85.115103}.

Eq.~(\ref{dual-Z}) is exactly same as the original lattice partition function
(see e.g. Eq.~(\ref{lattice})), no approximation is introduced 
either in the procedure of changing-variables or in the expansion of
the mixed term, as long as we carefully collect every term in
$V[d,d^{*}]$.   
However, due to the complexity of $V[d,d^{*}]$, we are partically
limited to a few lower order terms of it.
In most of the studies with the DF method, only the
two-particle reducible vertex in $V[d,d^{*}]$ is employed for calculating  
$G^{dual}_{k\omega_{n},\sigma}$. It turns to be sufficient of doing
so for constructing the non-local corrections to the DMFT solution in
most cases~\cite{PhysRevB.78.195105, PhysRevB.77.195105}.

\begin{figure*}[htbp]
\centering
\includegraphics[width=0.9\linewidth]{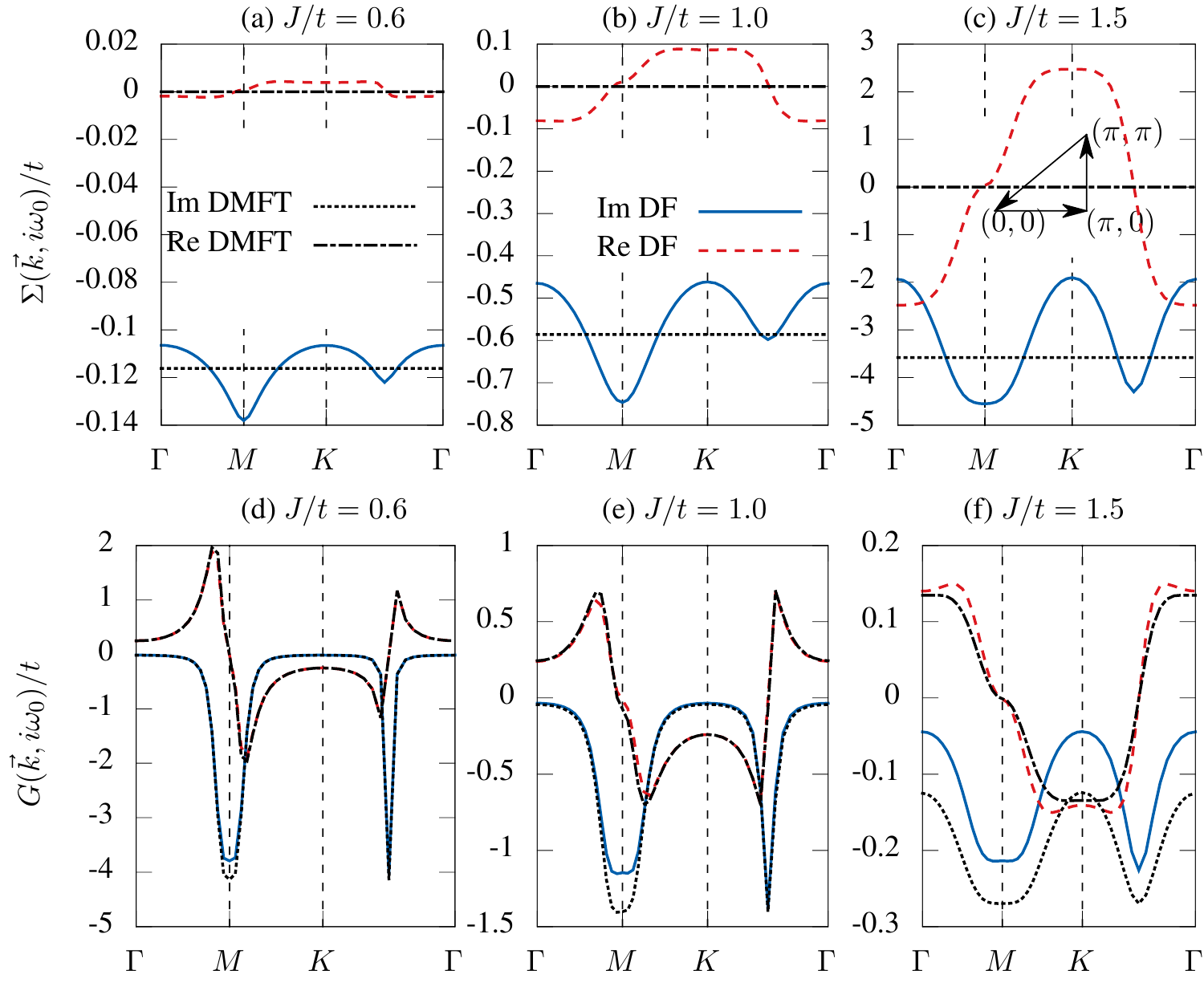}
\caption{The momentum evolution of the self-energy function of
  the anti-ferromagnetic Kondo model. The inverse temperature is
  set as $\beta t=25$. $\Sigma(\vec{k}, i\omega_{0})/t$ is shown for
  three different values of $J/t$ in (a), (b) and (c)
  respectively. The momentum dependence of $\Sigma(\vec{k},
  i\omega_{n})$ becomes more pronounced with the increase of $J/t$.} 
\label{SigmaK}
\end{figure*}

With the action of the dual variables,
\begin{equation}
S[d,d^{*}] =
\sum_{\alpha}d_{\alpha}^{*}G_{0,\alpha}^{dual,-1}d_{\alpha} + V[d^{*},d],
\end{equation}
one can perturbatively determine $G_{k\omega_{n},\sigma}^{dual}$ from
$G_{0,k\omega_{n},\sigma}^{dual}=1/[(\Delta(i\omega_{n})-\epsilon_{k})^{-1}
  + g_{\omega_{n},\sigma}]$
and $V[d, d^{*}]$. 
In terms of the two-particle reducible vertex, the self-energy
function of the dual variables can be calculated, {\it e.g.} from the
first two diagrams~\cite{PhysRevB.77.033101, PhysRevB.79.045133,
  PhysRevB.78.195105} in an expansion of $V[d, d^{*}]$, as 
\begin{eqnarray}
\Sigma^{dual}_{\alpha_{1}}&=&[\frac{T}{N}\sum_{2,3,4}G^{dual}_{\alpha_{3}}\gamma_{1234}
(\delta_{\alpha_{2};\alpha_{3}}-\delta_{\alpha_{1};\alpha_{2}})
\nonumber\\ 
&&-\frac{T^{2}}{2N^{2}}\sum_{2,3,4}G_{\alpha_{2}}^{dual}
G_{\alpha_{3}}^{dual}
G_{\alpha_{4}}^{dual}\gamma_{1234}\gamma_{4321}]\times\nonumber\\
&&\delta_{\alpha_{1}+\alpha_{3};\alpha_{2}+\alpha_{4}}.  
\end{eqnarray}
$G_{\alpha}^{dual}$ and $\Sigma_{\alpha}^{dual}$ are
related by Dyson equation $G_{\alpha}^{dual,
  -1}=G_{0,\alpha}^{dual, -1} -\Sigma_{\alpha}^{dual}$.
Through Eq.~(\ref{G-fd}), the frequency and
momentum dependent Green's function $G_{k\omega_{n},\sigma}$ can
straightforwardly be calculated. 

\subsection{results}\label{NL_Res}

The most convincing way of understanding the advantage of the
DF approach over the DMFT is to explicitly see the momentum
dependence of the self-energy function.
We show in Fig.~\ref{SigmaK} the evolution of $\Sigma(\vec{k}, \pi
t)/t$ as a function of $k$ along the path indicated in the middle of
Fig.~\ref{SigmaK}(c). 
These results correspond to the inverse temperature $\beta t=25$ and
only the self-energies for the lowest Matsubara frequency
$i\omega_{0}$ are shown here. 
As we know, in the DMFT local approximation, the impurity self-energy
contains only the dynamic information. It is a constant function of
$k$ in the 1st Brillouin Zone (BZ). 
In Fig.~\ref{SigmaK}(a-c), they are shown as straight lines in all
three plots, {\it i.e.} the DMFT self-energy has no dispersion
in momentum space. 
In the DF approach, due to the inclusion of the non-local
fluctuations, this dispersion is nicely restored. 
In Fig.~\ref{SigmaK}, the solid blue line and the dashed red line are
the corresponding imaginary and real parts of the self-energy, they
both display certain dispersions, which are missing in the DMFT
calculations.  
We find, this dispersion is much more pronounced for larger values
of $J/t$, indicating that in such case the DMFT local approximation
becomes less appropriate.
Interesting, we notice that the momentum dispersion is
always around the DMFT solution. 
Thus, a coarse-graining of the self-energy in momentum space would
lead to a constant value that is close to the DMFT solution. 
In this respect, we justify the DMFT to be essentially a reasonable
approximation to the problems in which electronic correlations
and the delocalization of electrons competes, as in the KLM.

\begin{figure}[htbp]
  \centering
  \includegraphics[width=\linewidth]{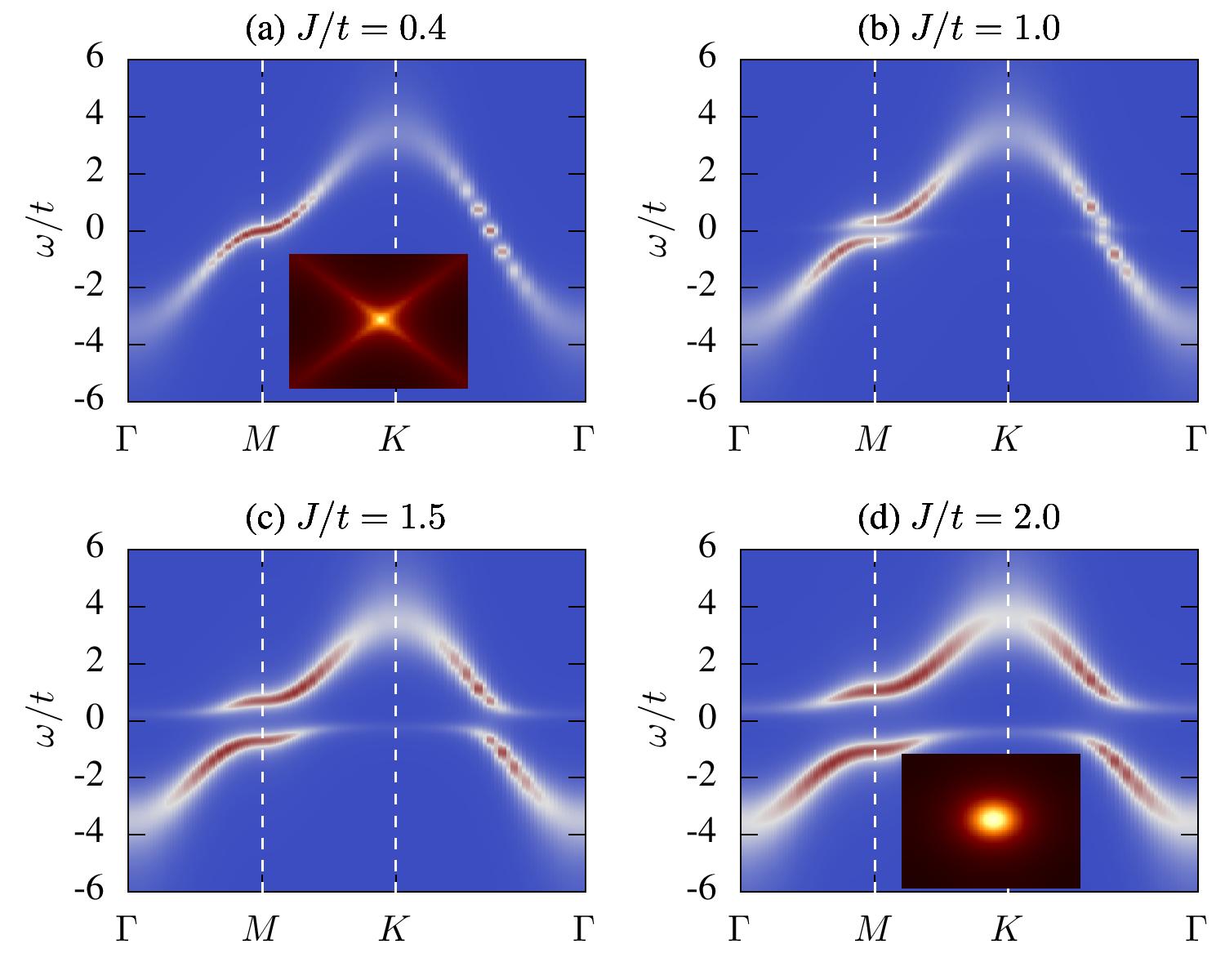}
  \caption{Single-particle spectral function of the conduction
    electrons at $\beta t=25$ for (a) $J/t=0.4$, (b) $J/t=1.0$, 
    (c) $J/t=1.5$ and (d) $J/t=2.0$. 
  The inset of (a) and (d) display the corresponding spin
  susceptibilities. At $J/t=1.0$ and $J/t=1.5$, the Kondo lattice
  model develops long-range anti-ferromagnetic order, thus spin
  susceptibilities are divergent and cannot be plotted.}
  \label{Akw}
\end{figure}
The deviation of $\Sigma(k,i\omega_{n})$ from the DMFT solution
$\Sigma(i\omega_{n})$ is more pronounced at $M$-point. With the
increase of $J/t$, the real-part of $\Sigma(M, i\omega_{n})$ becomes
flat around this point. 
As $M$-point is on the Fermi surface ($\epsilon_{M}=0$) of the
non-interacting conduction electrons, the non-local
corrections to 
$G(\vec{k},i\omega_{n})=1/(i\omega_{n}+\mu-\epsilon_{k}-\Sigma(\vec{k},
i\omega_{n}))$ from $\Sigma(\vec{k},i\omega_{n})$ would be much larger
here than at other values of $\vec{k}$. 
In Fig.~\ref{SigmaK}(d-f), the single-particle Green's function
$G(\vec{k}, i\omega_{n})$ are shown for the same parameters
corresponding to Fig.~\ref{SigmaK}(a-c).
As we explained above, $G(\vec{k}, i\omega_{n})$ from the DF
and the DMFT calculations are mainly different at $M$-point and
their difference becomes larger for larger $J/t$. 
For the same reason, though $\Sigma(\vec{k}, i\omega_{n})$ for
$\vec{k}=\Gamma$ also clearly deviates from the DMFT solution, $G(\vec{k},
i\omega_{n})$ gets less affected from it, since $\epsilon_{k}$ takes
the largest negative value at $\Gamma$-point. 
Only when the deviation becomes comparable with the half band-width,
{\it e.g.} as in Fig.~\ref{SigmaK}(c), $G(\Gamma, i\omega_{n})$ becomes
different from the DMFT one, see Fig.~\ref{SigmaK}(f).

What is more interesting is, that $\Sigma(\vec{k}, i\omega_{n})$
shows an additional symmetry which cannot be described by either
$\Gamma$, $M$ or $K$. 
In $K-\Gamma$ direction, we clearly observe another symmetry
line. 
The real-part of $\Sigma(\vec{k}, i\omega_{n})$ is anti-symmetric and
the imaginary-part is symmetric with respect to this symmetry. 
This is exactly where the Fermi surface of the non-interacting system
locates, and it is also the magnetic zone boundary for the $(\pi,
\pi)$-antiferromagnetic order.
Clearly, due to the inclusion of non-locality, the DF 
calculations capture the signature of the spin excitation. 
As the DMFT self-energy does not couple to any $k$-dependent
collective excitation, it is not possible for the DMFT to get this
additional symmetry.

Fig.~\ref{Akw} displays the single-particle spectral function
$A(\vec{k},\omega) = -\mbox{Im}G(\vec{k}, \omega)/\pi$ of the
conduction electrons with four different values of $J/t$ at $\beta
t=25$ in the paramagnetic phase. $A(\vec{k}, \omega)$ is obtained from
the corresponding single-particle Matsubara Green's function
$G(\vec{k}, i\omega_{n})$ from the stochastic analytic
continuation~\cite{Beach}.   
From Fig.~\ref{Gt} and Fig.~\ref{SE}, we learn that the ground state
of the anti-ferromagnetic Kondo model is insulating, at any $J/t$. 
At high temperature, we notice that the system can be a metal with
Fermi surface connecting the node $(\pm\pi/2, \pm\pi/2)$ 
and antinode $(0, \pm\pi), (\pm\pi, 0)$.
For low $J/t$, the RKKY interaction is expected to stabilize the
anti-ferromagnetic long-range order. 
As the appearance of the additional symmetry in
Fig.~\ref{SigmaK}(a-c), the inclusion of the $\vec{k}$-dependent
collective excitation in the DF approach is then expected
to resolve some precursor effects near the magnetic transition in
$A(\vec{k},\omega)$, {\it i.e. shadow band}~\cite{PhysRevB.82.245105}. 
Unfortunately, we did not observe any {\it shadow band} above/below
the Fermi level at $\Gamma/K$. 
This is probably because we simply took a constant error estimation in
the stochastic analytic continuation. 
{\it Shadow band} contains less spectral weight compared to the other
bands, which might be smeared out in the analytic continuation. 

\begin{figure}[htbp]
  \centering
\includegraphics[width=\linewidth]{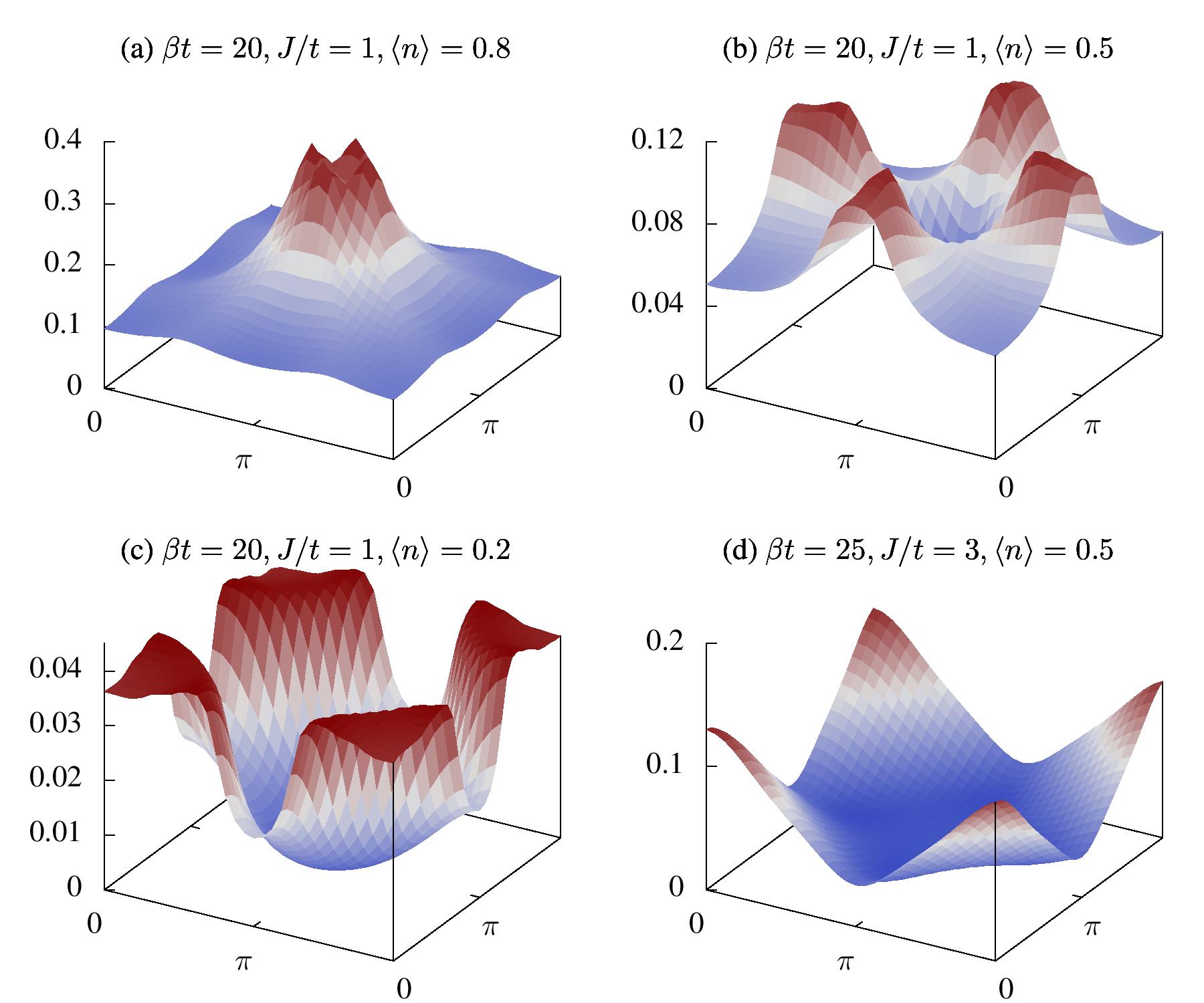}
\caption{The spin susceptibilities of the anti-ferromagnetic Kondo
  model at $J/t=0.5$ (upper row) and $J/t=1.0$ (lower row) for three
  different hole doping levels. 
A change from anti-ferromagnetic to ferromagnetic correlations is
observed with the hole doping, however, no stable ferromagnetic phase
is established for the cases studied here.}
\label{sus-nJ}
\end{figure}

\begin{figure*}[htbp]
\centering
\includegraphics[width=0.9\textwidth]{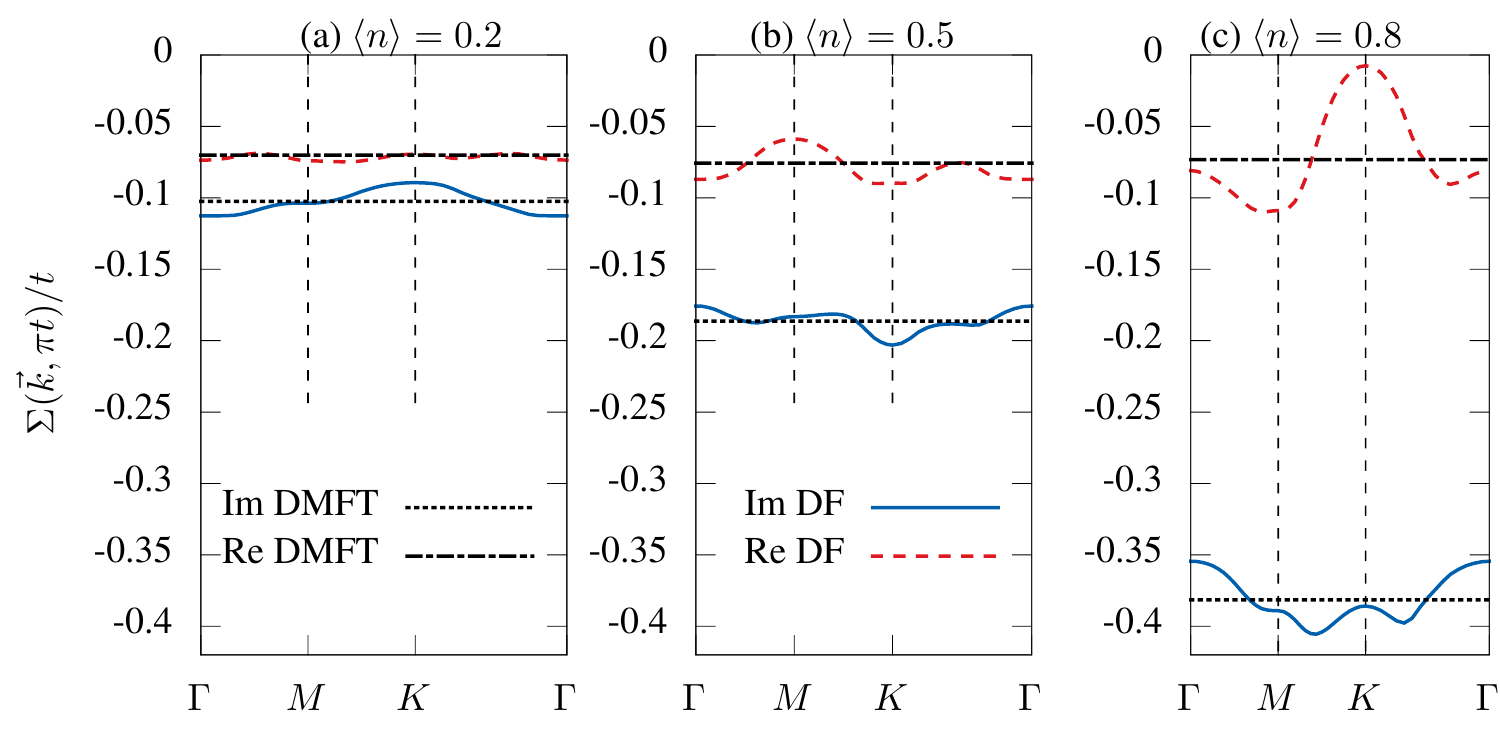}
\caption{The momentum dependence of the self-energy function at
  $J/t=1.0$ with different hole concentrations. 
Doping the system away from half-filling, the self-energy becomes flat
and the DMFT approximation becomes more justified.}
\label{SigmaK-nJ}
\end{figure*}

In addition to inducing {\it shadow bands}, the effective RKKY
interaction induced anti-ferromagnetic correlations can also drive the
system to insulating.    
At $J/t=1.0$ (see Fig.~\ref{Akw}(b)), the anti-ferromagnetic
long-range correlation is established and splits the bands
simultaneously at the Fermi level along the node and
antinode direction. 
Fig.~\ref{Akw}(a) and Fig.~\ref{Akw}(b), thus, show the metal-insulator
transition driven by the effective RKKY interactions and the
resulting anti-ferromagnetic correlations. 
When the coupling of the conduction electrons and the local moments
becomes larger, the Kondo effect starts to take effect. 
As shown in Fig.~\ref{Akw}(c), the valence band extends to $K$ from
both node and antinode. 
We would expect to have a large Fermi surface when we dope the system
with holes. 
At the same time, the conduction band extends to $\Gamma$ from node
and antinode. 
Similarly, if we dope the system with electrons, a small Fermi surface
will be obtained. 
Though, $A(\vec{k},\omega)$ in Fig.~\ref{Akw}(b, c, d) all contain a
charge gap at the Fermi level, the mechanism for gaping is essentially
different. 
In Fig.~\ref{Akw}(b), the conduction electrons is decoupled with the
local moments, the charge degree of freedom is gaped by the collective
spin excitation induced by the effective RKKY interaction.
In Fig.~\ref{Akw}(d), the Kondo singlet forms, the conduction electron
is bounded with local moments and, thus, is localized. 
Each site is occupied by one conduction electron
anti-ferromagnetically coupled with the local magnetic moments, while 
the empty and doubly-occupied states at half-filling only appear
virtually. 
The situation in Fig.~\ref{Akw}(c) is more complicated, thus, actually
more interesting. 
It is an anti-ferromagnetic insulator, the RKKY interaction is
still playing roles, however, the additional band features at $K$ and
$\Gamma$ are also apparent. 
Thus, the Kondo screening coexists with the anti-ferromagnetic
long-range order in this phase~\cite{PhysRevB.62.76, PhysRevB.63.155114}. 

The phase shown in Fig.~\ref{Akw}(d) is termed as
spin-liquid~\cite{RevModPhys.69.809}. 
The spin-liquid phase extends to $J/t=0$ in one-dimension Kondo
model~\cite{PhysRevLett.65.3177, PhysRevB.44.7486, PhysRevLett.71.3866}. 
At two-dimension, the RKKY interaction favors the anti-ferromagnetic
alignment of the spins between neighboring sites for small $J/t$,
the ground state is an anti-ferromagnetic insulator. 
The spin-liquid phase is favored only at high $J/t$, where the Kondo
interaction overwhelms the RKKY interaction. 
At high temperatures, the anti-ferromagnetic insulating states is
unstable with respect to the thermal fluctuations, while the Kondo
insulating states is found to be more robust. 
As a result, the ``metal"-``anti-ferromagnetic insulator"-``Kondo
insulator" transition is observed in Fig.~\ref{Akw} (Fig.~\ref{Akw}(a)
paramagnetic metal; Fig.~\ref{Akw}(b, c) antiferromagnetic insulator;
Fig.~\ref{Akw}(d) Kondo insulator).

The spin-liquid phase is magnetically disordered with the
corresponding spin susceptibilities being finite.
In the inset of Fig.~\ref{Akw}(d), the spin susceptibility
$\chi_{s}(Q)$ is shown in the entire 1st BZ.
$\chi_{s}(Q)$ peaks at $\vec{Q}=(\pi,\pi)$ and is rotationally
invariant about $(\pi,\pi)$. 
This is similar as the spin susceptibilities in the Hubbard model with
a small on-site Coulomb replusion $U$.
However, we need to note that in the Hubbard model, increasing $U$
leads to the enhancement of $\chi_{s}(Q)$. While, in the Kondo
model, further increase $J/t$ would suppress $\chi^{spin}(Q)$ as the
Kondo singlet would have lesser overlap with the neighboring ones.
Interestingly, in the paramagnetic metallic phase, {\it e.g.}
see Fig.~\ref{Akw}(a), the spin susceptibility shows a different
structure. 
In addition to the peak at $Q=(\pi, \pi)$,
$\chi_{s}(Q)$ also shows considerable amount of weight along $k_{x}=\pm k_{y}$.
The different structure of the spin susceptibility signals the
difference between the two paramagnetic states at low and high $J/t$
in terms of magnetic correlations, which is in line with the RKKY and
Kondo interactions.
In the RKKY regime, the electron spin susceptibilities are mainly
contributed by the electron-hole excitations along the perfect nested
Fermi surface.  
Due to the square shape of the Fermi surface in the 1st BZ, the
majority of the momentum vectors connecting two different pieces of
Fermi surface follows $k_{x}=\pm k_{y}$.
While, in the Kondo regime, the Fermi surface disappear, the low energy
excitation is between the top of the valence band and the bottom of
the conduction band. 
These portions of the bands in the 1st BZ are not straight
lines any more, but more extended. Thus, every combination of $k_{x},
k_{y}$ is possible, which results in the rotationally invariance of
the spin susceptibilities.

Now, let us move away from half-filling and study the destruction of
the spin liquid phase against hole doping.
When doping the system with holes, the Fermi energy moves into the
valence band, the spin-liquid phase evolves into the heavy Fermi
liquid state with enhanced quasiparticle mass and a large Fermi
surface. 
Intuitive mean-field studies of the phase diagram~\cite{10.1007}
indicate that, at two-dimension, for low $J/t$ the RKKY
anti-ferromagnetic phase extends to $\langle n\rangle\sim0.58$, then
it is replaced by a RKKY ferromagnetic phase. For large $J/t$, the Kondo
paramagnetic phase survives even with large hole concentrations,
only at very large $J/t$ the Nagaoka ferromagnetism overwhelms the
Kondo paramgnetism.
Similarly, the ferromagnetism at small $J/t$ in the one-dimension
Kondo lattice model was also numerically studied by many different
approaches, {\it e.g.} quantum Monte Carlo~\cite{PhysRevB.47.2886},
density matrix renormalization group~\cite{PhysRevB.65.052410}, the
DMFT~\cite{PhysRevB.86.165107}, exact
diagonalization~\cite{PhysRevB.77.073103}, etc. 
The ferromagnetic phase is found to be stable at filling $\langle
n\rangle < J/3t$. 
In addition, a ferromagnetic phase is also observed in the
intermediate filling region, which entirely embeds inside the
paramagnetic regime. 
Compared to the one-dimension case, less numerical results are known
for two-dimension Kondo lattice model at away-filling case, especially
for the magnetic ordering.  
Recently, Robert~\cite{PhysRevB.87.165133} and
Takahiro~\cite{PhysRevLett.110.246401} {\it etc.} found a charge ordered
phase at quarter-filling, which coexists with the anti-ferromagnetic
phase at low temperature.

Here, we study the magnetic correlations of the KLM with the DMFT and DF approaches.  
Our strategy here is to calculate the spin susceptibility in the
paramagnetic phase and to look for the divergence of it, which
corresponds to the instability of the paramagnetic solution. 
From which $Q$ the spin susceptibility $\chi_{s}(Q)$ becomes
divergent, we can interpret the type of the magnetic ordering that
breaks the paramagnetic symmetry. 
Fig.~\ref{sus-nJ} displays the spin susceptibilities for two values of
$J/t$ and three different hole concentrations. 
From $(a)\rightarrow(c)$, we see the peak of the spin
susceptibility at $(\pi, \pi)$ splits into four peaks and gradually
moves their position from $(\pi, \pi)$ to $(0, \pm\pi)$ and $(\pm\pi,
0)$. 
At the mean time, its amplitude is suppressed. 
Further increasing the hole concentration results in a continuous
movement of the peaks to $Q=\Gamma$, see Fig.~\ref{sus-nJ}(c), which
shows a tendency towards a ferromagnetic phase.
We find, the susceptibility basically does not change upon increasing
$J/t$ to moderate value like $J/t=1.0$ (not shown here) 
Only the structure gets smoother due to the same reason as for
Fig.~\ref{Akw}(a, d) (insets). 
However, further increasing $J/t$ and decreasing temperature $T$ will
induce a divergence in $\chi_{s}(\Gamma)$, see Fig.~\ref{sus-nJ}(d),
which possibly corresponds to the Nagaoka ferromagnetism. 
While, for small $J/t$, the ferromagnetic states are not really stable,
no long-range order could be detected in our calculations.
We further find, that the spin susceptibilities calculated from the
DMFT and the DF approaches agree with each other very well,
especially in the large hole doped case. 
Thus, we believe that the non-local fluctuation is weak in the large
doping regime.  
This can be further understood from Fig.~\ref{SigmaK-nJ}, where the
momentum dependence of the self-energy functions are shown for the
same doppings with $J/t=1.0$.
Close to half-filling, the non-local fluctuation strongly modifies the
self-energy function which induces an momentum dispersion as shown in
Fig.~\ref{SigmaK-nJ}(c).
Increasing hole concentration, the difference between the DMFT solution and
the DF solutions becomes less obvious.
The DF results now become closer to the
DMFT local solution, leading to the conclusion that the non-local
fluctuation becomes unimportant in the hole doped case, thus, the DMFT
should be a very reasonable and reliable approximation in this case.

\begin{figure}[htbp]
\centering
\includegraphics[width=\linewidth]{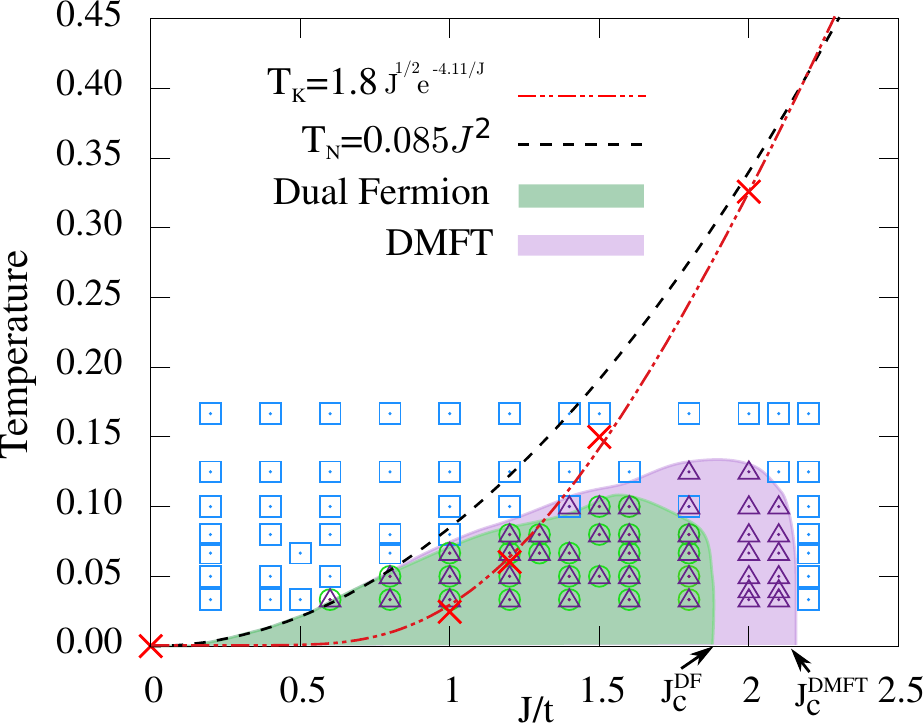}
\caption{Doniach diagram, where the antiferromagnetic regime and the
  Kondo insulator regime are separated at $J_{c}$. $J_{c}$ shows a
  sizable non-locality dependence from the comparison of
  $J_{c}^{DMFT}$ and $J_{c}^{DF}$. The light green and light purple
  color-filled regions are determined for corresponding
  antiferromagnetism in the DMFT and the DF, they only serves as the
  guide to the eyes. The exact solutions are given as symbols (see
  text for more details.)} 
\label{Mag}
\end{figure}
However, at half-filling, with respect to the size of the
antiferromagnetic phase, the DMFT shows a strong deviation from the
DF results. 
In Fig.~\ref{Mag}, we summarize our results for the antiferromagnetism
to the Kondo insulator transition, which is also discussed in
Fig.~\ref{Akw}, from the DMFT and the DF calculations. 
As stated before, by calculating the exact local two-particle vertex
from the CT-HYB, we can determine the spin susceptibilities
$\chi^{DMFT}_{s}(Q)$ and $\chi^{DF}_{s}(Q)$.
The antiferromagnetic phase are given by the divergence of
$\chi^{DMFT}_{s}(Q)$ and $\chi^{DF}_{s}(Q)$ at $Q=(\pi, \pi)$. 
In Fig.~\ref{Mag}, the paramagnetic solutions are shown as blue
square. 
In the DMFT, the antiferromagnetic phase are labeled as light-purple
triangles, which clearly shows different T-J
relations at large and small values of $J/t$. This highlights the
competition of the RKKY and the Kondo interactions. 
When $J/t$ is small, the N\'eel temperature $T_{N}$ increases with the
increase of $J/t$. 
$T_{N}$ can be fitted approximately as $T_{N}=0.085J^{2}$.
The same $J^{2}$-dependence of $T_{N}$ is also found in our DF
calculations, {\it i.e.} for smaller $J/t$, the DMFT and the
DF gives essentially the same solutions. 
At large $J/t$, in both the DMFT and the DF calculations, we find a quick
reduction of the N\'eel temperature, which drops to zero very rapidly
at $J_{c}^{DMFT}\sim2.15\pm0.05$ and $J_{c}^{DF}\sim1.85\pm0.05$.
The different values of $J_{c}$ in the DMFT and the DF calculations indicate that, the
non-locality becomes more relevant at larger $J/t$ side, which seems
to be unexpected.
As the RKKY interaction (the effective interaction between different
magnetic moments) in the smaller $J/t$ regime is non-local, while the
Kondo singlet is formed locally from the magnetic 
impurity and the polarization of the conduction electrons, we would
expect the non-local effect is more important in the RKKY regime, thus
the deviation of the DMFT and the DF to be larger at smaller
$J/t$. 
We believe this is due to the perfect nesting of the fermi surface at
smaller $J/t$, compared to which the non-locality becomes less
important and can be neglected. 
However, with the increase of $J/t$, the nesting fermi-surface is gradually
lost, non-local fluctuations become visible to the system.
Thus, the fermi-surface nesting suppresses the influence from the
non-local fluctuations,  
which leads to the nice agreement of the DMFT and the DF at smaller
$J/t$.
And for larger $J/t$, the non-local fluctuation effect starts to play
roles.

In Fig.~\ref{Mag}, the Kondo temperature is approximately determined
from the inverse of the impurity susceptibility from the DMFT, {\it i.e.}
$T_{K}=C_{Z}/\chi^{imp}_{s}(T)$, at $T/t=0.01$.
Following Doniach's energy argument, we find $T_{K}$ can be fitted as
$4.65 Exp[-5.26/J]$. 
By taking into account the spin-degeneracy, as suggested by
P. Coleman, we find $T_{K}$ can also be fitted approximately as
$T_{K}=1.8J^{1/2} Exp[-4.11/J]$. 
These two fittings are nearly on top of each other for the $J/t$
values studied here. 
The anti-ferromagnetic phase extends to $J/t=0$ limit in Fig.~\ref{Mag}, which partly due to the nested structure of the Fermi-surface at two-dimension. Therefore, it would be very interesting to know how the Doniach's diagram is modified when the Fermi-surface nesting is removed, for example, by the geometric frustration in a triangular system.~\cite{GangANDStefan}

\section{Conclusion}\label{Con}

In this paper, we present a detailed implementation of the dual
fermion approach, in conjugation with the hybridization expansion of
the continuous-time quantum Monte Carlo algorithm.
The single-particle Green's function and the two-particle reducible
vertex are sampled directly in the Matsubara frequency space, with
supplemented self-energy high-frequency tail. 
We applied our implementation to the two-dimension Kondo lattice
model, where we find that the non-local fluctuations is strong at
half-filling and becomes less important when the system is doped. 
The anti-ferromagnetic correlation induces one additional symmetry to
the self-energy function, which can be nicely explained by the N\'eel
magnetic ordering. 
At finite-temperature, we find a ``metal"-``anti-ferromagnetic
insulator"-``Kondo insulator" transition, which is resulted by the
competition of the effective RKKY interaction at low $J/t$ and the
Kondo effect at large $J/t$. 
The formation of the Kondo singlet opens the charge gap and induces
additional spectra around $K$ below the Fermi level and around
$\Gamma$ above the Fermi level.
Correspondingly, the change of the Fermi surface shape leads to
different structures in the spin susceptibilities. 
We confirm the Kondo effect and the anti-ferromagnetic correlations
coexist at half-filling.
We find the non-local fluctuations have more pronounced influence at the larger $J/t$ regime, 
the critical value of $J_{c}$ for the antiferromagnetism to the Kondo insulator transition is largely reduced when the non-locality is included.  
By doping the system, we further find that the anti-ferromagnetic
correlation is destroyed and the spin susceptibility tends to peak at
$Q=\Gamma$, which favors the ferromagnetic phase. 
However, no long-range ferromagnetic states is stabilized in our
calculations for small $J/t$.
In the doped case, the DMFT local approximation becomes very reliable
due to the fade of the non-local fluctuations in this case. 

\acknowledgments
The author  is grateful to the fruitful collaborations on the DF approach with A. N. Rubtsov, A.I. Lichtenstein, H. Monien, H. Lee, H. Hafermann, M.I. Katsnelson, S. Kirchner and A. Antipov in the previous projects.
Especially, the author wants to thank S. Kirchner for valuable discussions and W. Hanke for generous support while the preparation of the manuscript. 
The author thanks F. Assaad for providing the initial code for
preforming the stochastic analytic continuation.
This work is financially supported by the DPG Grant Unit FOR1162.

\bibliographystyle{apsrev4-1} 
\bibliography{ref}
 


\end{document}